\documentclass[twocolumn,aps,pra,showpacs,amsfonts,amssymb,amsmath,a4paper,flequ]{revtex4-1}
\usepackage{mathrsfs}
\usepackage{amsmath}
\usepackage{subfigure}
\usepackage{float}
\usepackage[title,titletoc,header]{appendix}
\usepackage{graphicx}

\begin{document}

 \draft
 \title{The confinement induced resonance in spin-orbit coupled cold atoms with Raman coupling}
 \author{Yi-Cai Zhang, Shu-Wei Song,  and Wu-Ming Liu}
 \address{Beijing National Laboratory for Condensed Matter
 Physics, Institute of Physics, Chinese Academy of Sciences, Beijing
 100190, China\nonumber }
\date{\today}

\begin{abstract}
 We investigate the confinement induced resonance in spin-orbit coupled cold atoms with Raman coupling. We find that the quasi-bound levels induced by the spin-orbit coupling and Raman coupling result in the Feshbach-type resonances.
  For sufficiently large Raman coupling, the bound states in one dimension exist only for sufficiently strong attractive interaction. Furthermore, the bound states in quasi-one dimension exist only for sufficient large ratio of the length scale of confinement
to three dimensional s-wave scattering length.
  The Raman coupling substantially changes the confinement-induced resonance position.  We give a proposal to realize confinement induced resonance by increasing the Raman coupling strength in experiments.

\end{abstract}

\maketitle

\emph{Introduction.---}For the low-dimensional quantum gas, the two-body scattering properties can be affected greatly by the external confinement potential. For example, when the s-wave scattering length is comparable to the transverse confinement length ($a_{3d}/a_\bot=1/C$ with $C=-\zeta(1/2)\approx1.46$), there exists a resonance, wherein the one dimensional effective interaction strength diverges \cite{Olshanii,Bergeman}. The similar scenario of the confinement induced resonance (CIR) also occurs in the qusi-two dimensional case \cite{Petrov}. The CIR has been observed through producing confinement induced molecules in qusi-one dimensional Fermi gas  \cite{Moritz}. It is also found experimentally that a single resonance splits into two resonances by introducing the anisotropic confinements \cite{Haller}. The transversally anisotropic confinement alters the position of resonance  by tuning the anisotropic ratio \cite{Zhang,Peng}. The CIR which can be used to tune the interaction between atoms, provides a crucial ingredient to realize the strong interacting low dimension systems,  such as Tonks-Girardeau gas \cite{Kinoshita,Paredes} and possible Tomonaga-Luttinger liquid \cite{Recati,Kakashvili}.

 The spin-orbit coupling (SOC) in cold atoms has attracted great interests in recent years.  The SOC could bring in nontrivial ground states in Bose-Einstein Condensate (BEC), such as  vortex or vortex lattice states, plane wave phase, standing wave phase \cite{Wu1,Zhai,Ho,Yip,Hu,Santos,Zhou,Guo,Zhou1}. The prominent effect induced by the SOC in fermions is that the SOC could enhance the low energy density of states, which results in the formation of two-body bound states and enhancement of the pairing gap \cite{Vyasanakere,Zhai2}.
In the presence of SOC, the two-body scattering properties in three dimension have been investigated. It is shown that the SOC usually results in the mixed-partial-wave scattering \cite{Cui}. For the low-energy scattering, the short range behaviors of wave function in three dimension can be modified by the SOC \cite{Zhangpeng,Yu}.  The two-body scattering properties in qusi-two-dimensional confinement with pure Rashba spin-orbit coupling are investigated \cite{zhangpeng}.

Some novel quantum states, for example, topological superfluidity, Majorana edge states or non-Abelian anyons could emerge in  the low dimension spin-orbit coupled quantum gas with Zeeman field \cite{Hu1, Sato,Wu}. In experiments, an effective Zeeman field in spin-orbit coupled atomic gas can be produced by two-photon Raman coupling \cite{Lin,Lin1,Wang2,Cheuk,Dalibard}. The Raman coupling strength corresponds the effective Zeeman field strength.
 The combination of the Raman coupling and SOC play an essential role in the formation of the above novel quantum states. The effects induced by the Raman coupling and SOC  are usually considered within BCS mean field framework \cite{Liu,Wei}.
  It is known that the two-body interaction properties provide basis for understanding the many-body system. The studies on the effects of the Raman coupling on two-body problem may give some insight into exotic quantum states.
   The CIR provides the indispensable tool for the realization of the low-dimensional strongly interacting quantum gas. Furthermore, how the Raman coupling and SOC affect CIR  is an inevitable question to clarify.
In the present Letter, we try to address the above questions through studying the two-body scattering problem in one dimension and the confinement induced resonance in the presence of the  Raman coupling and SOC.

\emph{Two-body scattering.--- }
We consider the Hamiltonian of spin-orbit coupled cold atoms with Raman coupling
\begin{align}
& H=H_K+H_0+V(x), \notag \\
& H_K=\frac{K^2}{4m}+\frac{\gamma K}{2m}(\sigma^x_2+\sigma^x_1), \notag \\
&H_{0}=\frac{k^2}{m}+\frac{\gamma k}{m}(\sigma^x_2-\sigma^x_1)+\frac{\Omega}{2}(\sigma^z_2+\sigma^z_1),
\end{align}
where the $H_K$ is the  Hamiltonian in center of mass coordinate of two atoms, $H_0$ is the free Hamiltonian in relative coordinate. $K$ and $k$ are total momentum and relative momentum of two atoms, respectively. $\sigma^x$ and $\sigma^z$ are the spin Pauli matrix, $V(x)$ is the interaction between two particles. $\gamma$ denotes the SOC strength and $\Omega$ is the two-photon Raman coupling strength between two Zeeman sublevels in experiment.  The SOC strength is determined by $\gamma=2\pi\hbar sin(\theta/2)/\lambda $, where $\lambda$ is the Raman laser wave length, $m$ is the mass of atom, $\theta$ is the angle between two Raman beams. The above Hamiltonian is realized experimentally in fermion atomic gas of $^{40}$K \cite{Wang2}. We take the natural units $m=1$, $\hbar=1$ and $\gamma=1$ in this section.

From Eq. (1), we get
that the motion of center of mass is coupled to the relative motion through spins.  In the following, we only focus on the subspace of Hamiltonian at $K=0$.
In the spin basis $[|s\rangle=(\uparrow\downarrow-\downarrow\uparrow)/\sqrt{2}, |t_1\rangle=\uparrow\uparrow, |t_2\rangle=\downarrow\downarrow, |t_3\rangle=(\uparrow\downarrow+\downarrow\uparrow)/\sqrt{2}]$, the Raman coupling and SOC are transformed into
$M\!\!=\!\!2\gamma[k/\sqrt{2}|s\rangle(\langle t_1|-\langle t_2|)\!\!+\!\!\emph{h.c}+\Omega/2\gamma(|t_1\rangle\langle t_1|-|t_2\rangle\langle t_2|)]$.
The interaction between cold atoms  can be modeled by zero-range pseudo potential.  Furthermore, we consider two identical spin-1/2 fermions. Hence,  only the s-wave  interaction in the singlet channel has contribution to two-body scattering.  Therefore the interaction matrix between two atoms takes the form as
$V(x)\!\!=\!\!|s\rangle\langle s| \!\otimes\! g_{1D}\delta(x)$,
where $g_{1D}$ is one dimensional interaction constant. From the matrix $V$ and $M$,  we know that the spin channel $|t_3\rangle$ is decoupled from other channels and not affected by the interaction. Thus, the spin channel $|t_3\rangle$ is dropped in the following, and the Hamiltonian $H_0$ is reduced to a 3 $\times$ 3 matrix. After diagnalizing $H_0$, the eigenenerges are obtained
$E_1(k)\!\!=\!\!k^2\!\!+\!\!2\gamma\sqrt{k^2+(\Omega/2\gamma)^2}$, $E_2(k)\!\!=\!\!k^2$, $E_3(k)\!\!=\!\!k^2\!\!-\!\!2\gamma\sqrt{k^2\!\!+\!\!(\Omega/2\gamma)^2}$ (see Fig. 1),  respectively. \begin{figure}[tbp]
\begin{center}
\includegraphics[ scale=0.5 ]{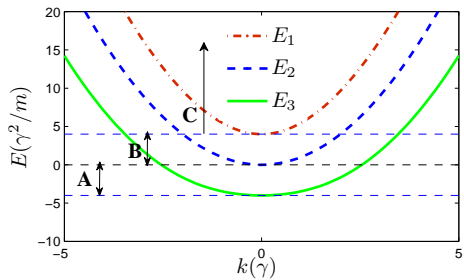}\\[0pt]
\end{center}
\caption{(Color online). The energy spectrum of the relative motion when the Raman parameter $\Omega=4$ (the SOC $\gamma=1$).  There are three energy branches, the highest $E_1$, the middle one $E_2$ and the lowest branches $E_3$.  In the case of $2\gamma< \Omega$, the lowest threshold is $-\Omega$. The A, B and C label three different scattering energy intervals $[-4,0]$, $[0,4]$ and $[4, \infty]$, respectively. The number of the scattering channels in different energy intervals are different.  }
\label{sw2d}
\end{figure}

\begin{figure}[tbp]
\begin{center}
\includegraphics[ scale=0.5 ]{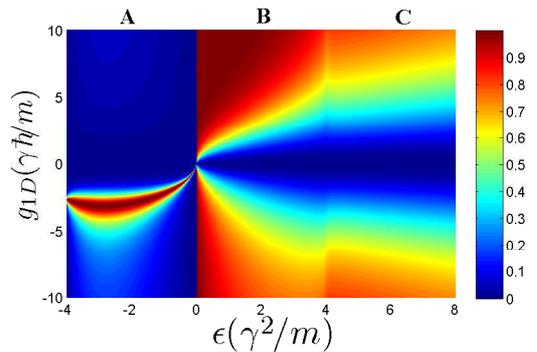}\\[0pt]
\end{center}
\caption{(Color online). The reflection coefficient $|f|^2$
 as a function of the interaction $g_{1D}$ and the scattering energy $\epsilon$ when $\Omega=4$ ($\gamma=1$).  The corresponding three scattering energy intervals in Fig. 1 are also labeled here.   }
\label{sw2d}
\end{figure}

 The scattering problem can be solved through the Lipmann-Shwindger equation
$\Psi(x)=\Psi_0(x)+\int dx^\prime G(\epsilon,x,x^\prime)V(x^\prime)\Psi(x^\prime)$,
where $G(\epsilon,x,x^\prime)=\sum_m\langle x|\frac{1}{\epsilon-E_m+i\eta}|x^\prime\rangle$ is the free Green function.
 In general, the Green function is $3\times3$ matrix. In the case of zero range interaction the Lipmann-Shwindger equation takes the following form

\begin{align}
\Psi(x)=\Psi_{0}(x)+ \frac{g_{1D}\Psi^1_{0}(0)}{1-g_{1D}G_{11}(0)}\left(                 
 \begin{array}{ccc}   
  G_{11}(x)    \\  
  G_{21}(x)   \\  
  G_{31}(x)
  \end{array}\right),
\end{align}
where $\Psi^1_{0}(x)$ is first component of incident state $\Psi_{0}(x)$.

It is known that in the usual case without Raman coupling and SOC, the reflection coefficient approaches one (total reflection) as the incident energy approaches the scattering threshold of $\epsilon=0$. For a fixed incident energy $\epsilon>0$, the reflection also approaches total reflection as the interaction $g_{1D}$ diverges. For attractive interaction $g_{1D}<0$, there always exist a bound state below the scattering threshold.

The scattering with Raman coupling and SOC
   is intrinsically multi-channel scattering problem \cite{Taylor}. There exist different scattering thresholds for different energy branches. When the incident energy crosses the thresholds, some scattering channels are opened or closed.
   In certain scattering energy interval, there may exist several scattering channels scattering each other (see Fig. 1).  The scattering amplitudes $f_{m,n}$ (reflection amplitudes) make up a matrix of rank 1.  It can be reduced to one single total amplitude $f$ by appropriately combining several scattering states in every energy interval.

 From Fig. 2, we can see that, comparing with the usual case, the Raman coupling and SOC cause fundamental changes in the behaviors of the scattering amplitude at low energy. First, there exist scattering resonances in the parameter space because there exist quasi-bound states between the energy branches \cite{Juzeliunas}. The interaction matrix $V$ can be rewritten in the eigen-basis of $H_0$. Besides the interaction in the respective eigen-basis channel (diagonal part), there are also non-diagonal part coupling to different eigen-basis. The energy branches above the incident energy could support bound states. In addition, the bound states are coupled to the scattering states due to the non-diagonal interaction term. Hence, it would result in a Feshbach-type resonance scattering \cite{Chin}.
Second, the reflection may vanish under certain conditions.  As shown in Fig.
 2, as the incident energy approaches the threshold of $\epsilon=0$ from below, the scattering amplitude becomes zero no matter how large the interaction is \cite{bu1}.
 Third, as the interaction $g_{1D}$ diverges, contrary to the usual case, the reflection need not mean a total reflection due to the effects of the closed channels in the other energy branches.
Four, the large Raman coupling changes the existing condition of bound states. When the Raman coupling satisfies $0<\Omega\leq 2\gamma$, there always exists a bound state below the lowest threshold for attractive interaction $g_{1D}<0$ as the usual case. However, if $\Omega>2\gamma>0$, there exists a bound state only when the attractive interaction is strong enough \cite{bu2}.

\emph{The confinement induced resonance.---}
  The one dimensional effective interaction constant is derived through investigating two-body problem of three dimension with confinement. The Raman coupling and SOC may change the condition of confinement induced resonance.
After separating the motion of center of mass, the Hamiltonian with confinement can be written as
\begin{align}
& H_r=H_0+H_\bot+V(r), \notag \\
& H_\bot=-\frac{\hbar^2}{2\mu}(\partial_{y}^2+\partial_{z}^2)+\frac{\mu \omega_{\bot}^2}{2}(y^2+z^2) ,
\end{align}
where $H_0$ is free Hamiltonian of relative coordinate along x direction as above section, $H_\bot$ is confinement potential, $V(r)=|s\rangle\langle s|\otimes g_{3D}\delta(\vec{r})\partial_r(r\cdot)$ is the three dimensional s-wave psuedo-potential interaction between atoms \cite{Huang}. $g_{3D}=4\pi \hbar^{2}a_{s}/m$ and $\omega_\bot$ are three dimension interaction constant and frequency of confinement, respectively. $ a_{s} $ and $\mu=m/2$ are  s-wave scattering length and reduced mass of two atoms, respectively. In this section, we take natural units as $m=1$, $\hbar=1$ and $\omega_\bot=1$.

In the following, we assume the confinement is strong enough that only the transverse harmonic ground state is occupied.
The incident energy with respect to the lowest threshold and the ground state energy of transverse harmonic oscillator should be  lower than the transverse exited state energy. The lowest excited state which can be coupled to ground states by the s-wave interaction is $\phi_1(y)\phi_1(z)$ with energy $2\hbar\omega_\bot$ \cite{Busch,Bergeman}. Hence, the incident energy should be smaller than $2\hbar\omega_\bot$.
 The wave function in three dimension is
$\Psi_{3D}(r)\!\!=\!\!\phi_0(y)\phi_0(z)\Psi_0(x)
\!\!+\!\!g_{3D}F\left(                 
  \begin{array}{ccc}   
    (G_{3D}(r))_{11}\\  
   (G_{3D}(r))_{21}\\  
    (G_{3D}(r))_{31}\\  
  \end{array}
\right)$,
where $\phi_n(t)\!\!\!=\!\!\!(\sqrt{\pi}2^{n} n!a_\bot)^{-1/2}e^{-t^2/2a^{2}_{\bot}}H_{n}(t/a_\bot)$, $a_\bot=\sqrt{\hbar/\mu \omega_\bot}=\sqrt{2}$ are  wave function and the length scale of Harmonic oscillator. $\Psi_0(x)$ is one dimensional incident wave function along x direction. $F=lim_{r\rightarrow 0}\partial_{r}[r (\Psi^{1}_{3D}(r)]$, $\Psi^{1}_{3D}(r)$ is the first component of the wave function. It is known that the three dimension wave function is singular at short range $\Psi(r)_{r\rightarrow 0}\propto 1/ r$. $F$ is the regular part of three dimension wave function at the $r=0$.

The resulting quasi-one dimensional wave function is
\begin{align}
&\Psi(\!x\!)\!\!=\!\!\Psi_0(\!x\!)\!\!+\!\!\frac{g_{3D}\Psi_{0}^{1}(0)}{1-g_{3D}(G_{3d}(0)_{11})_{r}}\!\!\! \left(\!\!\!                 
  \begin{array}{ccc}   
    (G_{3D}(x,0,0))_{11}\\  
   (G_{3D}(x,0,0))_{21}\\  
    (G_{3D}(x,0,0))_{31}\\  
  \end{array}\!\!\!
\right),\!\!\!
\end{align}
where $(G_{3D}(0)_{11})_{r}$ is the regular part of $(G_{3D})_{11}$ at the origin.
For scattering states, the long-ranged asymptotic behavior
 of three dimensional Green function behaves as $G_{3D}(x\!\!\gg\!\!1,0,0)\approx |\phi(0)|^4 G(x)$.
 The regular part is $(G_{3D}(0)_{11})_{r}\!\!\!\!=\!\!\!\!|\phi(0)|^4 G_{11}(0)+\sqrt{2}(C_1+C_2)/8\pi $, with
$C_1\!\!\!\!\!=\!\!\frac{-1}{\sqrt{\pi}}\int_{0}^{\infty}dt [\frac{e^{\epsilon t/2}}{(e^t-1)\sqrt{t}}-\frac{1}{t^{3/2}}]$,
$C_2\!\!=\!\!\frac{-1}{\sqrt{2}\pi}\int_{0}^{\infty}dt \frac{e^{\epsilon t/2}}{e^t-1} \int_{-\infty}^{\infty}dk \frac{ e^{-k^2 t/2}k^2[\texttt{cosh}(\gamma t \sqrt{k^2+(\Omega/2\gamma)^2})-1]}{k^2+(\Omega/2\gamma)^2}$.
 Comparing Eq. (4) with Eq. (2), one dimensional effective interaction constant is (restoring its units )
\begin{align}
g_{1D}&=\frac{2\hbar^2 a_{s}}{\mu a_{\bot}^2}\frac{1}{1-(C_1+C_2)a_{s}/a_{\bot}}.
\end{align}
This equation is our key result in this section, showing the connection between the one dimensional effective interaction constant and the three dimensional s-wave interaction constant.
When the SOC vanishes ($\gamma=0$), the $C_2$ becomes zero.
Furthermore, if the scattering energy $\epsilon \rightarrow 0$, the constant $C_1=C=-\zeta(1/2)\approx1.46$,
the resonance condition is reduced to Olshanii' result \cite{Olshanii}.

\begin{figure}[t]
\begin{center}
\includegraphics[ scale=0.60 ]{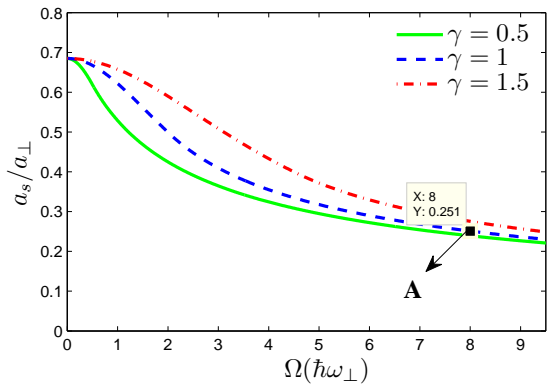}\\[0pt]
\end{center}
\caption{(Color online). The resonance position $a_{s}/a_\bot$ as a function of the Raman parameter $\Omega$ for different SOC strength $\gamma$. The point (\textbf{A}) denotes the resonance position with $\gamma=1$ and $\Omega=8$, which is consistent with the resonance position in Fig. 4 (see the red line).}
\label{sw2d}
\end{figure}
\begin{figure}[thp]
\begin{center}
\includegraphics[ scale=0.60 ]{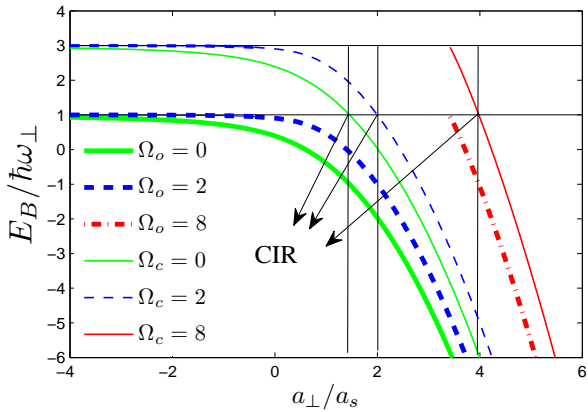}\\[0pt]
\end{center}
\caption{(Color online). The bound state energies in the closed channels and the full Hamiltonian, respectively (with SOC strength $\gamma=1$). The bound state energies are measured with respect to the lowest thresholds for different Raman parameter $\Omega$. The thin lines denote the bound state energies supported by the closed channels ($\Omega=\Omega_c$), and the thick lines are the bound state energies in the full Hamiltonian ($\Omega=\Omega_o$). The arrows denotes the positions of CIR. With the increase of Raman parameter $\Omega$, the resonance position $a_s/a_\bot$ (the reciprocal of $a_\bot/a_s$) is getting smaller and smaller, which is consistent with that in Fig. 3.}
\label{sw2d}
\end{figure}

When the Raman coupling is absent ($\Omega=0$) and the incident energy approaches the lowest threshold, the resonance condition is also exactly the same as the case without SOC ($a_s/a_\bot=1/(C_1+C_2)=1/C\approx0.68$) (see Fig. 3). It is related to the fact that the constant gauge potential can be gauged away by applying a gauge transformation when the Raman coupling is absent.  However, in the presence of non-zero Raman coupling strength, the resonance condition at the lowest threshold could never recover the usual case.
From Fig. 3, we can find that, with the increase of the Raman coupling, the resonance position $a_s/a_\bot$ is getting smaller. In addition, for sufficient large Raman coupling ($\Omega\gg1$), the resonance position incline to be independent of SOC strength. In fact, for fixed SOC parameter ($\gamma\sim1$), the resonance position $a_s/a_\bot \sim 1/\sqrt{2\Omega}$ can be arbitrary small by increasing the Raman parameter $\Omega$ \cite{bu3}.
It means that in the presence of the Raman coupling and SOC, the condition of CIR can be easier to fulfill than the usual case. For sufficiently large Raman coupling, the experimental observation of CIR  need not resort to the Feshbach resonance.

For fixed SOC strength $\gamma=1$, we show the bound state energies supported by the ``closed" channels (the transversely excited modes) and the full Hamiltonian, respectively in Fig. 4 \cite{Peng,Bergeman}.
The CIR can also be viewed as a Feshbach resonance as the usual  case without Raman coupling and SOC. The resonance
condition is satisfied when the energy of the bound states in closed
channel coincides with
the scattering threshold of the
ground transverse modes. Meantime the difference between the bound state energies in the closed channels and the full Hamiltonian is $2\hbar\omega_\bot$.
 However, the Raman coupling and SOC may change the existing condition of the bound states in the quasi-one dimension. When the Raman coupling satisfies $\Omega\leq 2\gamma$, there always exist the bound states irrespective of the sign of the s-wave scattering length $a_s$, which is consistent to the usual case \cite{Moritz} (also see the green and blue lines). However, if $\Omega>2\gamma$, the bound states exist only for sufficient large $a_\bot/a_s$ (see red lines).

\emph{The experimental realizations.---}
 The Raman coupling and SOC have been realized experimentally in $^{40}$K atomic gas \cite{Wang2}. Two magnetic sublevels $|\!\!\uparrow\rangle=|9/2,9/2\rangle$ and $|\!\!\downarrow\rangle=|9/2,7/2\rangle$ are chosen as two spin $1/2$ states. One can choose the experimental parameters $\gamma/m\sim 2\pi\hbar/(m\lambda)\approx1.28 cm/s$, $\omega_{\bot}\sim2\pi\times 17 \texttt{kHz}$ with $a_\bot\sim 172 nm$. The s-wave background scattering length $a_s\sim170 a_0\sim9nm$.
Under the above conditions, the SOC strength $\gamma\sim \sqrt{\hbar m \omega_\bot}$, the ratio $a_s/a_\bot=0.052$. To achieve the condition of CIR, one can increase Raman coupling strength ($\Omega\sim h\times3\texttt{MHz}$) by tuning the intensity of Raman beams.
The binding energies of bound states near CIR ($E_B\sim2\hbar\omega_\bot$) can be measured by using radio-frequency (rf) spectroscopy \cite{Regal0}. It is expected that there are two peaks in the radio-frequency photodissociation spectra. One peak locates  at the atomic transition frequency ($\nu_0$) of an occupied state to another initially unoccupied state (e.g. $|9/2,7/2\rangle\rightarrow |9/2,5/2\rangle$). The other peak at a non-zero detuning ($\delta=\nu_{rf}-\nu_0\sim34\texttt{kHz}$) from the atomic transition corresponds the dissociation of  quasi-one dimensional bound states \cite{Moritz}.

\emph{Conclusion.---}
In summary, we investigate the effects of the Raman coupling on the confinement induced resonance in the spin-orbit coupled cold atoms. The Raman coupling and spin-orbit coupling fundamentally change the interacting properties at low energy.  We propose to realize the confinement induced resonance through increasing Raman coupling strength. Different from the usual way, such as utilizing Feshbach resonances to produce a large scattering length, our work gives an new way to realize the strongly interacting quasi-one
dimension atomic gas with Raman coupling and spin-orbit
coupling.  Due to exotic effects induced by Raman coupling and spin-orbit coupling, a lot of  interesting many-body physical phenomena, e.g. the crossover of BCS- to BEC-like superfluidity \cite{Fuchs,Tokatly}, inhomogeneous Fulde-Ferrell-Larkin-Ovhinnikov (FFLO) state \cite{Hu2},  fermion pair breaking in presence external magnetic field \cite{Guan}, need to be revised in the strong interacting quasi-one dimension atomic gas.

\noindent{\bf Acknowledgements:}
We thank Lin Wen, Fadi Sun, Fei Zhou, Ran Qi, Daw-Wei Wang and Xiaolin Cui for useful discussion. This work was supported by the NKBRSFC under grants Nos. 2011CB921502, 2012CB821305, 2010CB922904, and NSFC under grants Nos. 61227902 and NSFC-RGC under grants Nos. 11061160490 and 1386-N-HKU748/10.


\begin{references}






\bibitem{Olshanii} M. Olshanii, Phys. Rev. Lett. \textbf{81}, 938 (1998).


\bibitem{Bergeman} T. Bergeman, M. G. Moore, and M. Olshanii, Phys. Rev. Lett. \textbf{91}, 163201 (2003).

\bibitem{Petrov} D. S. Petrov and G. V. Shlyapnikov, Phys. Rev. A \textbf{64}, 012706 (2001).





\bibitem{Moritz} H. Moritz, T. St\"{o}ferle, K. G\"{u}enter, M. K\"{o}hl, and T. Esslinger, Phys. Rev. Lett. \textbf{94}, 210401 (2005).


\bibitem{Haller} E. Haller, M. J. Mark, R. Hart, J. G. Danzl, L. Reichs\"{o}llner,  V.  Melezhik,
P. Schmelcher, and H. C. N\"{a}gerl, Phys. Rev. Lett. \textbf{104}, 153203 (2010).



\bibitem{Peng} S. G. Peng, S. S. Bohloul, X. J. Liu, H. Hu, and P. D. Drummond, Phys. Rev. A \textbf{82}, 063633 (2010).
\bibitem{Zhang} W. Zhang and  P.  Zhang, Phys. Rev. A \textbf{83}, 053615 (2011).



















\bibitem{Kinoshita} T. Kinoshita, T. Wenger, D. S. Weiss, Science \textbf{305}, 1125 (2004).




\bibitem{Paredes} B. Paredes, A. Widera, V. Murg, O. Mandel,
S. F\"{o} lling, I. Cirac, G. V. Shlyapnikov,
T. W. H\"{a}nsch and I. Bloch, Nature \textbf{429}, 277 (2004).


\bibitem{Recati}  A. Recati, P. O. Fedichev, W. Zwerger, and P. Zoller,  Phys. Rev. Lett. \textbf{90}, 020401 (2003).

\bibitem{Kakashvili} P. Kakashvili, S. G. Bhongale, H. Pu, C. J. Bolech,  Physica B \textbf{404}, 3320 (2009).










\bibitem{Wu1} C. J. Wu, I. M. Shem, and X. F. Zhou, Chin. Phys. Lett. 28,
097102 (2011).
\bibitem{Zhai} C. Wang, C. Gao, C. M. Jian, and H. Zhai,  Phys. Rev. Lett. \textbf{105},  160403 (2010).
\bibitem{Ho} T. L. Ho and S. Zhang,  Phys. Rev. Lett. \textbf{107},  150403 (2011).

\bibitem{Yip} S. K. Yip, Phys. Rev. A \textbf{83}, 043616 (2011).








\bibitem{Hu} H. Hu, B. Ramachandhran, H. Pu, and X. J. Liu,  Phys. Rev. Lett. \textbf{108},  010402 (2012).


\bibitem{Santos} S. Sinha, R. Nath, and L. Santos,  Phys. Rev. Lett. \textbf{107},  270401 (2011).

\bibitem{Zhou} X. F. Zhou, J. Zhou, and C. J. Wu, Phys. Rev. A \textbf{84},  063624  (2011).


\bibitem{Guo} S. W. Su, I. K. Liu, Y. C. Tsai, W. M. Liu, and S. C. Gou, Phys. Rev. A \textbf{86},  023601  (2012).

\bibitem{Zhou1} X. Zhou, Y. Li, Z. Cai, C. Wu,
J. Phys. B: At. Mol. Opt. Phys. \textbf{46} 134001 (2013)










\bibitem{Vyasanakere} J. P. Vyasanakere and V. B. Shenoy, Phys. Rev. B \textbf{83}, 094515 (2011).
\bibitem{Zhai2} Z. Q. Yu, and H. Zhai,  Phys. Rev. Lett. \textbf{107},  195305 (2011).



\bibitem{Cui} X. Cui, Phys. Rev. A \textbf{85}, 022705 (2012).
\bibitem{Zhangpeng} P. Zhang, L. Zhang, and Y. Deng, Phys. Rev. A \textbf{86}, 053608 (2012).
\bibitem{Yu} Y. Wu and Z. Yu, Phys. Rev. A \textbf{87}, 032703 (2013).

\bibitem{zhangpeng} P. Zhang, L. Zhang, W. Zhang, Phys. Rev. A \textbf{86}, 042707 (2012).






\bibitem{Hu1} H. Hu, L. Jiang, H. Pu, Y. Chen, and X. J. Liu,  Phys. Rev. Lett. \textbf{110}, 020401 (2013).


\bibitem{Sato} M. Sato, Y. Takahashi, and S. Fujimoto,  Phys. Rev. Lett. \textbf{103}, 020401 (2009).

\bibitem{Wu} F. Wu, G. C. Guo, W. Zhang, and W. Yi,  Phys. Rev. Lett. \textbf{110}, 110401 (2013).

\bibitem{Lin1} Y. J. Lin, K. J. Garc\'{\i}a and I. B. Spielman, Nature \textbf{471}, 83 (2011).
\bibitem{Lin} Y. J. Lin, R. L. Compton, A. R. Perry, W. D. Phillips, J. V. Porto, and I. B. Spielman,  Phys. Rev. Lett. \textbf{102},  130401 (2009).
\bibitem{Wang2} P. J. Wang, Z. Q. Yu, Z. K. Fu, J. Miao, L. H. Huang, S. J. Chai, H. Zhai and J. Zhang,  Phys. Rev. Lett. \textbf{109},  095301 (2012).
\bibitem{Cheuk} L. W. Cheuk, A. T. Sommer, Z. Hadzibabic, T. Yefsah, W. S. Bakr, and M. W. Zwierlein,  Phys. Rev. Lett. \textbf{109}, 095302 (2012).

 \bibitem{Dalibard} J. Dalibard, F. Gerbier, G. Juzeli\={u}nas and P. \"{O}hberg,  Rev. Mod. Phys. \textbf{83}, 1523 (2011).







\bibitem{Liu} X. J. Liu, Phys. Rev. A \textbf{87}, 013622 (2013).
\bibitem{Wei} R. Wei, E. J. Mueller, Phys. Rev. A \textbf{86}, 063604 (2012).

\bibitem{Taylor} J. R. Taylor, Scattering Theory (Wiley, New York, 1972).

\bibitem{Juzeliunas} A similar resonance is obtained in a spin-independent potential scattering problem of spin-orbit coupled atoms by J. J. Ruseckas, R. Jur\v{s}\.{e}nas, G. Juzeli\={u}nas and I.B. Spielman, unpublished.







\bibitem{Chin} C. Chin, R. Grimm, P. Julienne and E. Tiesinga, Rev. Mod. Phys. \textbf{82}, 1225 (2010).



\bibitem{bu1} This is because when the incident energy approaches the threshold of zero energy from below ($\epsilon \rightarrow 0_{-}$), the middle energy branch $E_2$ contributes a infinitely large real part ($G_{11}(0)\sim\frac{1}{2\pi}\int^\infty_{-\infty} dk\times [\frac{1}{\epsilon-E_2+i\eta}\frac{(\Omega/2)^2}{k^2+(\Omega/2)^2}] \approx 1/2\sqrt{-\epsilon}$) in the  denominator of the scattering amplitude (see Eq. (2)).
 \bibitem{bu2} This is because when the incident energy approaches the lowest threshold ($\epsilon\rightarrow -\Omega$ ), all the energy branches contribute a finite part ($G_{11}(0)=\frac{1}{2\pi}\int^\infty_{-\infty} dk\times [\frac{1}{\epsilon-E_2+i\eta}\frac{(\Omega/2)^2}{k^2+(\Omega/2)^2}
+\frac{1}{\epsilon+i\eta-E_1}\frac{k^2}{k^2+(\Omega/2)^2}+\frac{1}{\epsilon+i\eta-E_3}\frac{k^2}{k^2+(\Omega/2)^2}]$) in the denominator of the scattering amplitude,  rather than an infinitely large number as in usual case without Raman coupling and SOC.







\bibitem{Huang} K. Huang, and C. N. Yang, Phys. Rev. \textbf{105}, 767 (1957).

\bibitem{Busch} T. Busch, Be. G. Englert, K. Rz\c{a}\.{z}ewski, and M. Wilkens, Found. Phys. \textbf{28},  549 (1998).








\bibitem{bu3} It can be shown that, for a fixed spin-orbit coupling ($\gamma= 1$), the $C_1\sim\sqrt{2\Omega}$ diverges and the $C_2$ is bounded as  $\Omega\rightarrow\infty$.
 When the condition of CIR is satisfied, the resonance position  $a_s/a_\bot =1/(C_1+C_2)\sim1/\sqrt{2\Omega}$.


\bibitem{Regal0} C. A. Regal, C. Ticknor, J. L. Bohn and D. S. Jin, Nature \textbf{424}, 47 (2003).






\bibitem{Fuchs} J. N. Fuchs, A. Recati, and W. Zwerger,  Phys. Rev. Lett. \textbf{93}, 090408 (2004).



\bibitem{Tokatly} I.V. Tokatly,  Phys. Rev. Lett. \textbf{93}, 090405 (2004).

\bibitem{Hu2} H. Hu, X. J. Liu, and P. D. Drummond,  Phys. Rev. Lett. \textbf{98}, 070403 (2007).

\bibitem{Guan} X. W. Guan, M. T. Batchelor, C. Lee, and M. Bortz, Phys. Rev. B \textbf{76}, 085120 (2007).




\end{references}
\end{document}